\documentclass[pdflatex,sn-standardnature]{sn-jnl}%

\usepackage{amsmath,amsfonts,amssymb}

\begin{document}

\title{The Oxygen Bottleneck for Technospheres}

\author[1]{Amedeo Balbi}

\author[2]{Adam Frank}

\affil[1]{Dipartimento di Fisica, Università di Roma `Tor Vergata', I-00133 Roma, Italy}

\affil[2]{Department of Physics and Astronomy, University of Rochester, Rochester, NY 14620, USA}


\abstract{As oxygen is essential for respiration and metabolism for multicellular organisms on Earth, its presence may be crucial for the development of a complex biosphere on other planets. And because life itself, through photosynthesis, contributed to creating our oxygen-rich atmosphere, oxygen has long been considered as a possible biosignature. Here we consider the relationship between atmospheric oxygen and the development of technology. We argue that only planets with substantial oxygen partial pressure ($p_{\rm O_2}$) will be capable of developing advanced technospheres and hence technosignatures that we can detect. But open-air combustion (needed, for example, for metallurgy), is possible only in Earth-like atmospheres when $p_{\rm O_2}\ge 18\%$. This limit is higher than the one needed to sustain a complex biosphere and multicellular organisms. We further review other possible planetary atmospheric compositions and conclude that oxygen is the most likely candidate for the evolution of technological species. Thus, the presence of $p_{\rm O_2}\ge 18\%$ in exoplanet atmospheres may represent a contextual prior required for the planning and interpretation of technosignature searches.}

\keywords{Astrobiology, extrasolar planets, extraterrestrial intelligence, technosignatures}



\maketitle


Atmospheric oxygen (O$_2$) has played an important role in astrobiological studies. Because it is an essential component of respiration and metabolism for multicellular organisms on Earth, some researchers have argued that the presence of free oxygen in the atmosphere will be essential for the evolution of complex life and intelligence on exoplanets as well \cite{Knoll1985,Catling2005,Judson2017}.

What has not received much attention, though, is the role of O$_2$ in the development of technology on a planetary scale, or `technospheres', of the kind that `technosignature' studies are designed to find \cite{Socas-Navarro2021, Wright2022}. In particular, we might ask how does atmospheric chemistry limit or enhance the potential for a species which already has evolved tool-using intelligence to take the next step and begin developing higher order technological innovations.
In particular, we might ask how does atmospheric chemistry limit or enhance the potential for a species which has already evolved tool-using intelligence to take the next step and begin developing higher order technological innovations. If a global scale technological civilization cannot arise without a substantial amount of atmospheric O$_2$, then the limits on oxygen partial pressure ($p_{\rm O_2}$) set a bottleneck for the emergence of `technospheres' on other planets.

To our knowledge, this issue has not been addressed before. The purpose of this paper is to lay out the problem, offer some initial observations about specific constraints, and outline the methods that may be useful in determining answers.  

\section{Oxygen regulation in the atmosphere and biology}

\begin{figure}
\includegraphics[width=\textwidth]{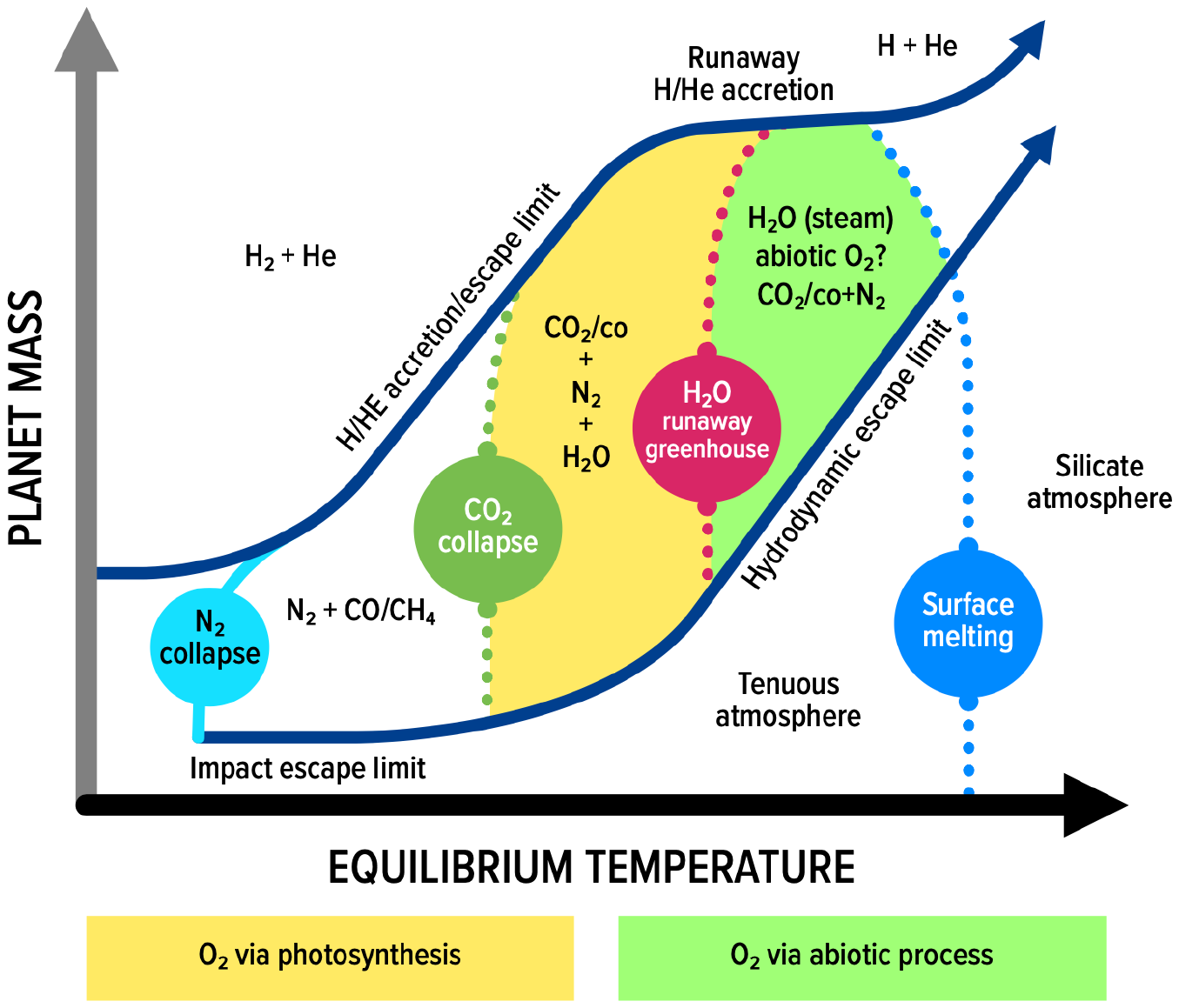} \\
\caption{{\bf Planets capable of supporting high O$_2$ concentrations and, hence, technological civilizations}.  This figure shows the likely composition of planetary atmospheres based on mass and equilibrium temperature. For planets whose temperature and mass would lead to primarily CO$_2$/N$_2$/H$_2$O atmospheres, high oxygen levels require a biological origin, i.e. photosynthesis.  For higher equilibrium temperatures, high O$_2$ levels can occur via the runaway greenhouse effect and the photolysis of water vapor at high altitudes. (Figure adapted with permission from \cite{Forget2014}, Royal Society Publishing.)}
\label{PlanetAtmo}
\end{figure}

The composition of a planet's atmosphere depends on many factors including its astronomical origins and geologic history (Figure~\ref{PlanetAtmo}). What transpired on Earth strongly suggests that the presence of considerable quantities of O$_2$ in the atmosphere over long periods of time probably requires the presence of a biosphere.

\begin{figure}
\includegraphics[width=\textwidth]{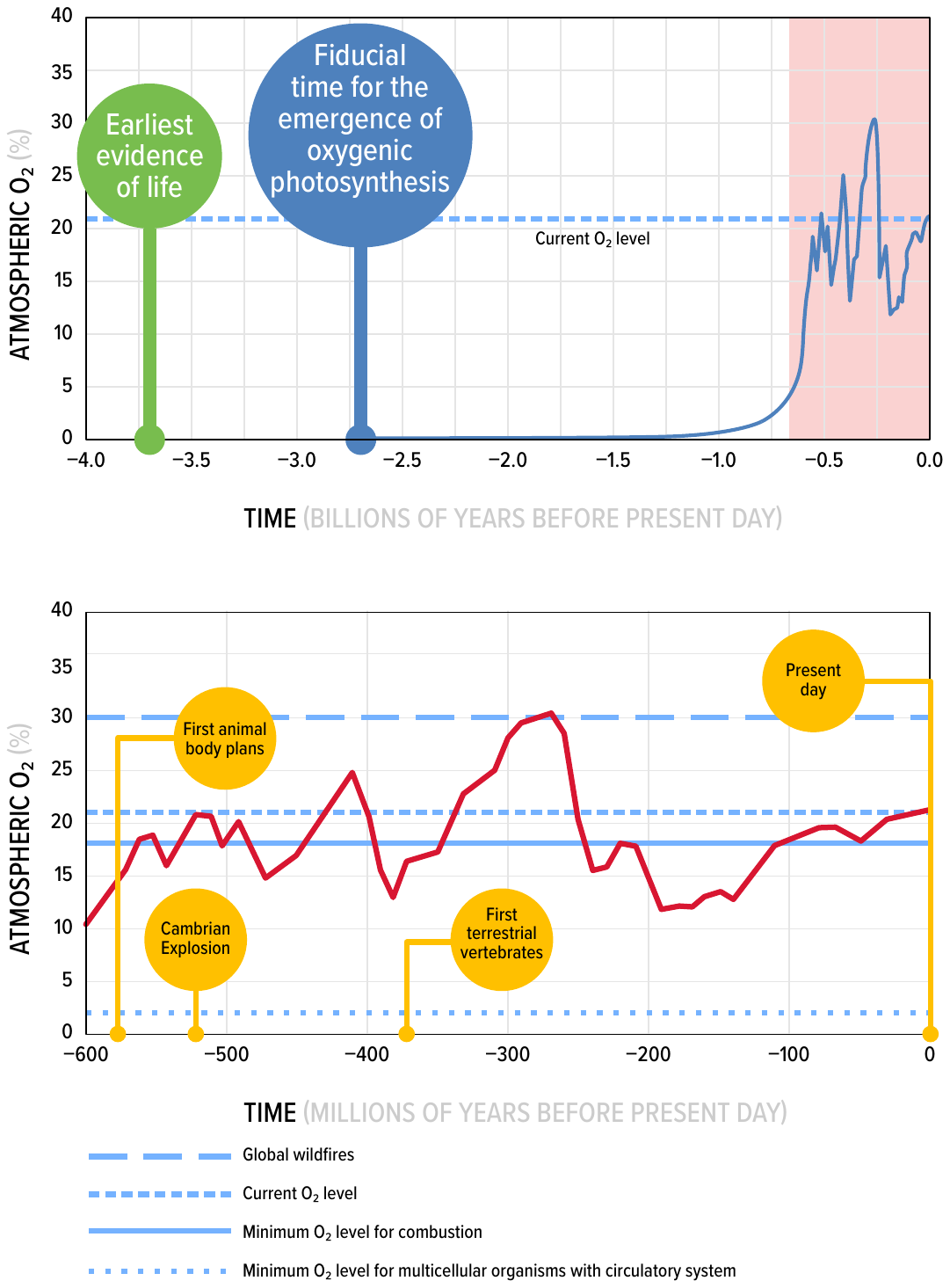} \\
\caption{{\bf Earth's atmospheric O$_2$ concentration over time.} Top: O$_2$ concentration beginning at 4 Gyr ago. The Great Oxidation Event around 2.4-2.2 Gyr ago (see text) is not apparent due to the choice of linear scale for the vertical axis. The shaded red region is enlarged in the bottom panel.  Bottom: O$_2$ concentrations beginning 600 Myr ago, around the development of multicellular life. The horizontal lines represent the current $p_{\rm O_2}$ level, the threshold for global wildfires, the minimum requirement for combustion and the minimum value for the smallest vascularized organisms.  In both panels we label important events in the evolution of life.   (Oxygen data from \cite{Berner2007}. Figure adapted with permission from \cite{Costa2014}, under a Creative Commons licence CC-BY 3.0..)}
\label{EarthO2}
\end{figure}

The history of oxygen in Earth’s atmosphere is a complicated one (Figure~\ref{EarthO2}). Broadly speaking, O$_2$ is primarily released by oxygenic photosynthesis, but since organic carbon is rapidly oxidized (either by aerobic respiration or by the decay of dead organisms), the net source of free O$_2$ is strongly dependent on the rate of carbon burial through the subduction of tectonic plates in the mantle. Also important are other processes that can deplete the atmosphere of O$_2$, such as reactions with reduced minerals or volcanic gases. These factors are not easy to constrain even on modern Earth, and it is hard to predict how they might act on other planets (see \cite{Catling2017} for a thorough discussion). However, it is generally believed that a substantial rise in atmospheric oxygen and its permanence over geological times requires the evolution of photosynthetic organisms as a necessary precondition.

Although the actual mechanisms for the rise of O$_2$ concentration in Earth’s atmosphere are still debated, there is strong evidence for two pivotal events \cite{Rye1998, Lyons2014}. The first, termed Great Oxidation Event (GOE), occurred around 2.4-2.2 Gyr ago, and resulted in a rapid increase of the O$_2$  volume concentration from less than 1 part per million to 0.06-2\%. This is consistent with the timing for the appearance of the first photosynthetic organisms, which is estimated around 2.7 Gyr ago and possibly earlier \cite{Buick2008, Catling2020}. The second happened around 750-520 Myr ago, when the O$_2$ concentration rose again well above 3\%, until it reached the present value of 21\% by volume. Besides these noteworthy variations, O$_2$ levels fluctuated considerably during the last 550 million years (the Phanerozoic eon) \citep{Berner2006, Berner2009,Krause2022}  (Figure \ref{EarthO2}).

The interplay between biology and atmospheric O$_2$ works in both directions, as a high level of free O$_2$ might have been a necessary precondition for the appearance of complex life \cite{Catling2005}. Aerobic metabolism is about ten times more energy effective than its anaerobic counterparts for a given food intake. Also, basic chemistry suggests that no other element can do better than oxygen as an energy source in biology. Only fluorine and chlorine are more energetic oxidants (per electron) than oxygen, but they both have drawbacks: they are not as abundant as oxygen in the universe, and they have undesired side effects in organic reactions at moderate temperatures (fluorine easily explodes, while chlorine in aqueous environment produces compounds that destroy organic molecules). Also, both F$_2$ and Cl$_2$ cannot be abundant in gas form in planetary atmospheres due to their high reactivity. Therefore, free molecular oxygen is plausibly the most energetic source available to life in any planetary environment (note, however, that there are counter-arguments for fluorine-based life: see, e.g., \cite{Budisa2014}). 

Indeed, high levels of O$_2$ have been linked to the emergence of multicellularity \cite{Knoll2017} and to the appearance and growth of animal life \cite{Planavsky2014}. While single eukariotic cells or small multicellular organisms are possible at O$_2$ concentrations as low as 0.2\%, O$_2$ levels must be at least 2\% to reach the minimal size needed for basic vascularization and circulation, $\sim 1$ mm \cite{Catling2005}. Values as high as  $\sim 12$\% are probably needed for animal size comparable to the smallest mammals existing today, $\sim 3 $ cm \cite{Catling2005, Lingam2021}.

Size considerations are especially important for the development of intelligent life, requiring large and complex brains. It has also been noted \cite{Catling2005, Went1968} that the use of sophisticated technology might be impossible for organisms below a minimal size. This would create an indirect link between O$_2$ levels and the rise of technological species. 

There is, however, a more direct connection: technology requires combustion, and this sets strict limits on the minimal amount of free atmospheric O$_2$ available. 

\section{Physics and Chemistry of Combustion}

On Earth, combustion can occur in open air environments because of the presence of a fairly high $p_{\rm O_2}$.  But combustion, as a generic form of chemical reaction, does not require such O$_2$  atmosphere.  Combustion reactions are generally defined as requiring a `fuel' compound and an `oxidant'.  These are reacted to produce heat and a new product. Combustion reactions often require an activation energy barrier to be crossed (i.e.\ lighting a match) but once initiated they are exothermic, with enough energy liberated to be self-sustaining, if the flux of fuel and oxidizer is sufficient.  

Typical reactions readily available on Earth are
\begin{eqnarray}
& & \rm 2 H_2 + O_2 \rightarrow 2H_2O + heat\\
& & \rm CH_4 + 2 O_2 \rightarrow CO_2 + 2 H_2O + heat
\end{eqnarray}

Note that in these reactions the oxidant is oxygen (O$_2$ in both cases).  But this need not be the case.  Combustion is an electron transfer redox reaction that moves electrons (or electron density) from a less to a more `electronegative' atom.  This transfer will be both spontaneous and will release energy, which is why combustion, if possible, may be the first energy source a young intelligent species learns to master.  While the transfer of electrons from fuel to oxidant defines combustion, the oxidant does not have to be oxygen.  Other elements and compounds can play the same role role in combustion.  Fluorine and chlorine are both possible examples of oxidants.  Fluorine combustion involves some of the most rapid and exothermic reactions known 
\cite{Jones1987} and has been proposed as an oxidizer for extreme high-performance propulsion systems.  This is due to the fact that fluorine is the most electronegative element known. Other possibilities include N$_2$O, a strong oxidizing agent which is known to drive combustion of hydrocarbons \citep{Razus2022}. All such alternative combustion mechanisms, however, are problematic in Earth-like conditions and result in explosive reactions that are difficult to control or self-sustain.

It is worth noting that combustion using oxygen as an oxidant can also occur without the presence of appreciable oxygen in an atmosphere.  In the modern Earth atmosphere, fires can burn in open air because $p_{\rm O_2}$ is high enough that it is the atmospheric oxygen molecules that serve as oxidant.  But in controlled settings the atmospheric O$_2$ levels can be reduced to the point where it stops being the source of oxidant and some other compound can supply the required oxygen atoms.  In these cases the presence of an oxidizer in some form of condensed states reduces the need for O$_2$ from the air.  Atmospheric O$_2$ may then still play a role in combustion but that role is secondary \cite{Biteau2010}.

\section{Oxygen and technology}

On Earth, fire certainly played a crucial role in the establishment of human civilization and the onset of technology. The date for the first controlled use of fire is uncertain, with some evidence setting it as early as 1.5 million years ago \cite{Gowlett2016,Hlubik2019}. There are several ways in which the use and control of fire gave substantial advantages to our species. Cooked food yields higher energy density compared to its raw counterpart, and its consumption arguably produced physiological changes (larger brains, smaller teeth and digestive apparatus; see, e.g., \cite{Zink2016}) as well as lasting social effects \cite{Wrangham2010}. Fire is also essential as a tool to change the environment: foragers may have been using it to aid hunting and to control vegetation growth as early as 55,000 years ago \cite{Mellars2006, Archibald2012}. However, by far the most important role for fire in the rise of human civilization is its use as an energy source. This probably started very early on as a way to keep warm, thereby increasing the range of environmental conditions suitable for settlements; then it evolved into more sophisticated uses, such as metal smelting, melding and tool fabrication;  finally, it provided both the energy source and the fuel (e.g.\ charcoal) that initiated the industrial revolution and led to the `Great Acceleration' and Anthropocene \cite{Steffen2015}. 

If we use the development of radio technologies as our standard of a technologically `advanced' intelligence that might be detected across interstellar distances, then our focus on fire (combustion) and oxygen may rest on the development of metallurgy.  All early complex civilizations on Earth relied on `worked' metals such as copper and tin that were mass produced \cite{Smil2017}.  Copper, in particular, appears in both weapons and tools as early as the 6th millennium BCE.  Such large-scale deployment of metals involves the application of heat in a variety of ways, the most basic of which is in smelting, which is the process by which the metal is extracted from raw ore. The creation of mixtures of metals, known as alloys, to enhance properties such as hardness or resistance to corrosion, also requires the application of heat.  Alloys are formed by melting combinations of the metals together and then mixing them before they are allowed to cool.  Copper’s relatively high melting point at 1083°C meant that it could not be purified from ores and then worked without the combustion of substantial quantities of wood or charcoal.  Bronze, which played a pivotal role in the development of premodern societies, is an alloy of copper and tin.  Since tin ore contains anywhere between 5\% and 30\% purity levels, the melting point required for its use ranged between 750 and 900°C.  Iron, and its alloy steel, eventually took over as the main metal for industrial production but it required even higher melting temperature of 1535°C.  

Thus the most important developments of metallurgy, i.e.\ the ones which led to our ability to create an industrial base that could produce something like a radio telescope, all relied on combustion to raise metal temperatures high enough so that phase changes from solid to liquid could be achieved.  From the first metal-working societies to the early industrial era, this was done via the combustion of biomass.  Fuel was supplied by wood (perhaps in the form of charcoal) and the oxidizer was atmospheric oxygen.  Thus we can next enquire what atmospheric O$_2$ levels are required for `open air' combustion.

Experimental studies using paper as fuel have shown that the lower limit of O$_2$ concentration for combustion in Earth’s atmosphere occurs around 16\%.  The probability of fire ignition, however, is strongly inhibited below 18.5\%, and it is only above a concentration of 20\% that ignition can be assured \cite{Belcher2008, Belcher2010}. It is noteworthy that the present-day O$_2$ concentration on Earth is 21\%, which seems to sit around the sweet spot for combustion.  It has been suggested that coupled Earth system feedbacks might help maintaining O$_2$ around this level, as concentrations above 25\% would increase the likelihood of widespread fires ignited by lightning and quickly suppress vegetation (and therefore photosyntesis) on the land surface, which would in turn cause a reduction in O$_2$ \cite{Lenton2000}. Other studies have challenged this idea, showing that, depending on moisture content, oxygen levels above 30\% might still allow for the persistence of terrestrial plants \cite{Wildman2004}.  Anyway, concentrations around 35\% would likely be the highest compatible with the existence of vegetation \cite{Chaloner1989}. 

O$_2$ levels are still important even when the atmospheric O$_2$ is not the primary oxidizer. Experimental studies \cite{Biteau2010} found that when O$_2$ concentration was reduced under 18\%, large variations were observed in the CO$_2$ and CO concentrations (the reaction products), suggesting that below this threshold the reactions occur only through the oxidizer contained in the material.  Thus, $p_{\rm O_2}\simeq 18\%$ seems to be the limit for both ignition and maintenance of combustion in open air atmospheric conditions.

Consideration of these findings implies that controlled use of fire on Earth (and, by extension, the development of technology) would have been problematic during periods when $p_{\rm O_2}< 18\%$, and probably not feasible at all for  $p_{\rm O_2}< 16\%$. Given the history of O$_2$ concentration, the flammability of Earth could have been highly variable even during the Phanerozoic, possibly switching off completely for a few tens of million years around 180 and 200 million years ago (see Figure~\ref{EarthO2}) \cite{Falkowski2005}. Had a tool-using species evolved during these periods, it would not have been able to cast metals into the forms needed to produce `advanced' technologies like radio telescopes. Generalizing this conclusion to other planets implies the existence of comparable lower limits to the atmospheric O$_2$ concentration required for the presence of a detectable technosphere.

Although we have focused on metallurgy, combustion is important for many other processes that have contributed to development of industrial societies.  The firing of bricks for construction is a very early example of the need for combustion beyond metallurgy.  Other modern examples include the Haber process, for the production of nitrogen based fertilizer, which optimized for production at 450°C, and the refinement of crude oil into petroleum which requires temperatures 370°C. Space-faring capabilities would also be indirectly dependant on the availability of oxygen, as in the liquid form this is the most commonly adopted oxidiser in rocket propellants. 

\section{Conclusions}

The main point of this article is to call attention to the fact that the oxygen concentration in a planetary atmosphere required for flammability is well above the minimal level needed to sustain a complex biosphere and multicellular organisms (Figure~\ref{EarthO2}). As combustion is an indispensable prerequisite for the onset of advanced forms of technology -- such as those required for communication over interstellar distances -- this would set constraints on the kind of planetary environment capable of maintaining a detectable technosphere. Such constraints would be far more stringent than those required for the presence of complex life and even intelligence.  

Various forms of alternative biochemistry can be conjectured for both simple life and (to some extent) complex life \cite{Budisa2014} relying on oxidants other than oxygen to harness chemical energy. In many cases these would, however, require unusual and problematic atmospheric chemistry, like high fluorine concentrations. It would be beyond the scope of this discussion to revisit all the alternative chemistries proposed for `exotic' life: if, however, we restrict ourselves to Earth-analogues as a conservative template for a habitable planet, O$_2$ seems the most plausible oxidant available for both efficient metabolism and combustion.  

The development of combustion-based technology on Earth-like planets, therefore, does seem to be heavily dependent on the availability of O$_2$ at high concentrations. Of course, speculative scenarios involving alternative ways to harness energy for technological use on other planets are in principle feasible. For example, a primitive extraterrestrial intelligent species could produce heat by focusing stellar light, by tapping geothermal energy and perhaps even by naturally occurring nuclear reactions. However, compared to such alternatives, combustion of biomass has many obvious advantages: most notably, it is globally available, self-sustainable, portable and relatively easy to master and reproduce. Such advantages have been noted elsewhere \citep{Smil2017} and arguably make combustion the most accessible and versatile means to kickstart the rapid progress of a young civilization. 

While the build-up of an appropriate O$_2$ concentration seems a requisite for the development of a technological civilization, this condition is not met automatically by the requirement that complex life is possible. Intelligent life can conceivably evolve in an environment with low availability of free O$_2$ (for example in water, or in an atmosphere with O$_2$ levels lower than present Earth, etc.)\cite{Lingam2023}, and ascend to the point of developing communicating capabilities and tool utilization, but it will eventually lack access to the large energy densities that are readily available through combustion. This would be a manifest hindrance for the development of advanced technology. Such a strong bottleneck was not taken into account in previous studies of technosignatures. Intriguingly, the same restriction would apply even for extreme forms of artificial or post-biological life that, although not directly dependent on oxygen, would still need it in sufficient quantity in the atmosphere for large-scale industrial activity to function.

The existence of an oxygen bottleneck has important implications for future searches of technological activities on exoplanets. Target selection might prioritize planets with spectral O$_2$ signatures that are above the combustion threshold and, at the same time, the presence of high O$_2$ levels (or lack thereof) should serve as a contextual prior to assess the credibility of possible evidences of technosignatures in the data. Also, the existence of a potentially direct connection between O$_2$ levels and technological capabilities may assist in estimating the frequency of advanced civilizations, using theoretical models of geochemical and biological processes that have an impact on atmospheric O$_2$ concentration (e.g.\ photosynthesis, tectonics, global element cycles, etc.). This might give some handle on one of the unknown factors appearing in the Drake equation \cite{Drake1965}, namely, $f_c$, the fraction of planets where intelligent life develops the technological capabilities required for interstellar communication. Should theoretical predictions show that high O$_2$ levels are not a likely output of planetary evolution, this would provide a partial solution to the so called `Fermi paradox', or the lack of evidence for intelligent life elsewhere. 

\backmatter

\bmhead{Acknowledgments}
We are grateful to David Catling and Manasvi Lingam for useful conversations.

\bmhead{Competing Interests}
The authors declare no competing interests.





\bibliography{biblio}


\end{document}